\newcommand{\nc}{\newcommand}
\nc{\be}{\begin{equation}}
\nc{\ee}{\end{equation}}
\nc{\bea}{\begin{eqnarray}}
\nc{\eea}{\end{eqnarray}}
\nc{\beas}{\begin{eqnarray*}}
\nc{\eeas}{\end{eqnarray*}}
\nc{\noi}{\noindent}
\nc{\sD}{\not \! \! D}
\nc{\s}[1]{\not \! \!  #1}
\nc{\non}{\nonumber}
\nc{\bb}{\bibitem}
\nc{\lf}{\left}
\nc{\ri}{\right}
\nc{\mb}[1]{\makebox[#1]{}}
\nc{\pa}{\partial}
\nc{\sA}{\not \! \! A}
\nc{\newsec}[1]{\section{#1}\mb{0.5cm}}
\nc{\ra}{\rightarrow}
\nc{\la}{\leftarrow}
\nc{\lapp}{\hbox{$ {     \lower.40ex\hbox{$<$}
                   \atop \raise.20ex\hbox{$\sim$}
                   }     $}  }
\nc{\rapp}{\hbox{$ {     \lower.40ex\hbox{$>$}
                   \atop \raise.20ex\hbox{$\sim$}
                   }     $}  }
\nc{\ep}{$e^+e^-\ra\pi^+\pi^-\;$}
\nc{\M}{{\cal M}}
\nc{\rhoom}{$\rho^0$-$\omega\;$}
\nc{\CP}{{\s{\rm CP}}}
\documentstyle[preprint,aps]{revtex}
\tighten
\begin{document}
\draft
\preprint{\vbox{                       \hfill UK/TP 97-09  \\
                                       \null \hfill ADP-97-15/T252 \\
                                       \null\hfill hep-ph/9705453 \\
}}

\title{Rho-Omega Mixing and Direct CP Violation \\
      in Hadronic B-Decays}
\author{S. Gardner and H.B. O'Connell}
\address{Department of Physics and Astronomy, \\ University of Kentucky, 
        Lexington, KY 40506-0055 USA}
\author{A.W. Thomas}
\address{Department of Physics and Mathematical Physics, \\ and 
Special Research Centre for
the Subatomic Structure of Matter, \\
University of Adelaide, 
Adelaide, S.A. 5005 AUSTRALIA}

\date{October 29, 1997}
\maketitle
 
\begin{abstract}
The extraction of CKM-matrix-element
information from hadronic B-decays generally
suffers from discrete ambiguities, hampering the diagnosis of physics
beyond the Standard Model.
We show that a measurement of the rate asymmetry, which is CP-violating, in 
$B^{\pm}\rightarrow\rho^{\pm}\rho^0(\omega)\rightarrow\rho^{\pm}\pi^+\pi^-$,
where the invariant mass of the $\pi^+\pi^-$ pair is
in the vicinity of the $\omega$ resonance, 
can remove the {\rm mod}($\pi$) uncertainty in 
$\alpha\equiv {\rm arg} [-V_{td} V_{tb}^\ast/(V_{ud}V_{ub}^\ast)]$ 
present in standard analyses.
\end{abstract}
\pacs{PACS numbers: 11.30.Er, 11.30.Hv, 12.15.Hh, 13.20.He }

\narrowtext

Although CP violation in the neutral kaon system has been known 
since 1964~\cite{flrev96}, 
it is not yet known whether the Cabbibo-Kobayashi-Maskawa (CKM) matrix, 
and hence the Standard Model, 
provides a correct
description of CP violation. The next generation of $B$-meson experiments 
will address both the 
empirical determination of the CKM 
matrix elements and the issue of 
whether a single CP-violating parameter, as in the Standard Model,
suffices to explain them. 
Hadronic decays of $B$-mesons will play an important role in elucidating the 
CKM matrix elements, and many clever methods have been devised to evade
the uncertainties the strong interaction would weigh on their 
extraction~\cite{flrev96}. 
Nevertheless, discrete ambiguities in the CKM matrix elements remain, for in 
$B^0-\bar{B}^0$ mixing the weak phase $\phi$ 
enters as $\sin 2\phi$~\cite{GQ97}. It is our purpose to demonstrate
that it is also possible to determine the sign of $\sin\phi$ through
the measurement of the rate asymmetry in
$B^{\pm}\rightarrow\rho^{\pm}\rho^0(\omega)\rightarrow\rho^{\pm}\pi^+\pi^-$,
where
the invariant mass of the
$\pi^+\pi^-$ pair is
in the $\rho^0$-$\omega$ interference region,
so that the ${\rm mod}(\pi)$ 
ambiguity consequent to the $\sin 2\phi$ measurement is removed. 
This is necessary to test 
the so-called unitarity triangle associated with the
CKM parameters $\alpha$, $\beta$, and $\gamma$, for the Standard
Model requires that these
angles sum to $\pi$~\cite{pdg96}.

In $B^{\pm}\rightarrow\rho^{\pm}\rho^0(\omega)\rightarrow\rho^{\pm}\pi^+\pi^-$
decays, proposed by Enomoto and Tanabashi~\cite{ET96}, CP violation is
generated by the interference between a $b\ra u$ tree amplitude and a 
$b \ra d$ penguin amplitude, or their charge conjugates. 
The rate asymmetry, which is CP-violating, arises exclusively from a 
nonzero phase in the CKM matrix, so that the CP violation is
termed ``direct.'' 
The latter's existence requires that at least two amplitudes
contribute and that both a strong and weak phase difference exists
between them~\cite{bss79}. 
If the internal top quark dominates 
the $b \ra d$ penguin amplitude, 
then the weak phase in the rate asymmetry 
enters as $\sin\alpha$, where 
$\alpha\equiv 
{\rm arg} [-V_{td} V_{tb}^\ast/(V_{ud}V_{ub}^\ast)]$~\cite{pdg96}. 
In $B^0-\bar{B}^0$ mixing, in contrast, $\sin 2\alpha$ enters. 
Strategies to determine this latter quantity include the study of 
$B^0 \ra \pi\pi$~\cite{gronau90}, 
$B^0 \ra \rho\pi$~\cite{snyder93},
$B^0 \ra \pi\pi$ and $B^0 \ra \pi K$~\cite{dighe96}, or the latter  
with $B_s^0$ decays as well~\cite{kim97}. The last two methods 
assume the top quark dominates the $b \ra d$ penguin, though 
the non-negligible charm quark mass implies that the 
GIM cancellation of the up and charm quark 
contributions to the $b \ra d$ penguin is not perfect~\cite{buras95}.  
However, as the $t$ quark penguin contribution is numerically 
larger~\cite{buras95}, the 
sign of $\sin\phi$ suffices to determine that of $\sin\alpha$. 
For our purposes, then, $\phi$ is proportional to $\alpha$. 

Grossman and Quinn have suggested that the $B\ra \pi\pi$ and
$B\ra\rho\pi$ analyses mentioned above 
can be combined to determine $\cos 2\alpha \sin\alpha$ and hence
$\sin\alpha$~\cite{GQ97}.
However, their analysis requires the use of the factorization approximation
to estimate the sign of a 
ratio of hadronic matrix elements -- the phase of this ratio
is the strong phase. 
In the factorization approximation, the 
hadronic matrix elements of the four-quark operators are 
{\it assumed} to be saturated by vacuum intermediate states. This 
approximation can be justified in QCD in the limit of a large number of
colors~\cite{buras86}, 
and it finds phenomenological justification in a comparison
with measured B-decay branching ratios~\cite{rodrig96}. 
We also use the factorization approximation, 
but the channel we propose has the important advantage that it permits
a significant test of its 
applicability. This, we believe, is unique to the channel we study and
is only possible because 
$e^+e^- \ra \pi^+ \pi^-$ data in the $\rho^0$-$\omega$ 
interference region provides additional 
hadronic information.

The CP-violating asymmetry in 
$B^{\pm}\rightarrow\rho^{\pm}\rho^0(\omega)\rightarrow\rho^{\pm}\pi^+\pi^-$
in the vicinity of the $\omega$ resonance 
is predicted to be more than 20\% that of 
the summed decay rates 
with a branching ratio ($B^\pm\ra \rho^\pm\rho^0$) 
of more than  $10^{-5}$~\cite{ET96,lipkin93,branch}. 
Interfering resonances can generally both
constrain and enhance the strong phase~\cite{atwood95}. 
Here we can also extract the non-resonant
strong phase; only its quadrant need be computed.
To understand why the asymmetry 
is significantly enhanced by $\rho^0$-$\omega$ 
mixing, consider the amplitude 
$A$ for $B^- \ra \rho^- \pi^+ \pi^-$ decay:
\be
A = \langle \pi^+\pi^- \rho^- | {\cal H}^{\rm T} | B^- \rangle
+ \langle \pi^+\pi^- \rho^- | {\cal H}^{\rm P} | B^- \rangle \;,
\label{start}
\ee
where ${\cal H}^{\rm T}$ and ${\cal H}^{\rm P}$ correspond to the tree
and penguin diagrams, respectively. Defining the strong phase
$\delta$, the weak phase $\phi$, 
and the magnitude $r$ via 
\be
A = \langle \pi^+\pi^- \rho^- | {\cal H}^{\rm T} | B^- \rangle \left[
1 + re^{i\delta}\; e^{i\phi} \right] \;,
\ee
one has 
$\overline{A}
=\langle \pi^+\pi^- \rho^+ | {\cal H}^{\rm T} | B^+ \rangle 
[1 + re^{i\delta}\; e^{-i\phi}]$. 
Here $\phi$ is $-\alpha$ if the top quark dominates
the $b\ra d$ penguin.
Thus, the CP-violating 
asymmetry, $A_{\CP}$, is 
\be
A_{\CP} \equiv 
{ | A |^2 - |{\overline A}|^2 
\over | A |^2 + |{\overline A}|^2 }
= {-2r \sin \delta \sin \phi 
\over 1 + 2r\cos \delta \cos \phi + r^2 } \; .
\label{asym}
\ee

If we are to calculate $A_{\CP}$ reliably and hence determine $\sin \phi$
we need to know the strong phase, $\delta$.
Let $t_{\rm V}$ be the tree 
and $p_{\rm V}$ be the penguin amplitude for 
producing a vector meson ${\rm V}$. 
In terms of the $\rho$ and $\omega$ propagators, $s_{\rm V}^{-1}$ 
(with $s_{\rm V}=s - m_{\rm V}^2 + i m_{\rm V} \Gamma_{\rm V}$ and 
$s$ the invariant mass of the $\pi^+ \pi^-$ pair), and the
effective $\rho - \omega$ mixing amplitude, $\tilde\Pi_{\rho\omega}$, we find: 
\bea
\langle \pi^+\pi^- \rho^- | {\cal H}^{\rm T} | B^- \rangle 
= {g_{\rho} \over s_\rho s_\omega} \tilde\Pi_{\rho\omega} t_{\omega}
  + { g_{\rho} \over s_\rho } t_\rho  \;, \;\;
\langle \pi^+\pi^- \rho^- | {\cal H}^{\rm P} | B^- \rangle 
= {g_{\rho} \over s_\rho s_\omega} \tilde\Pi_{\rho\omega} p_{\omega}
  + { g_{\rho} \over s_\rho } p_\rho \;, \label{fun2}
\eea
where $g_\rho$ is the $\rho^0 \ra \pi^+\pi^-$ coupling.
Maltman {\it et al.}~\cite{hoc97}
have considered the ``direct'' coupling $\omega \rightarrow \pi^+\pi^-$, 
not included in Ref.~\cite{ET96}, as well as the ``mixing'' 
contribution $\omega \ra \rho^0 \ra \pi^+\pi^-$. 
Fortunately, this additional term can be absorbed into an
energy-dependent, effective mixing amplitude, 
$\tilde\Pi_{\rho\omega}(s)$~\cite{pionff97}.
The latter has recently been extracted~\cite{pionff97} from 
the world data for the reaction
$e^+e^-\ra \pi^+\pi^-$ \cite{barkov85}, for $\sqrt s$ near the $\omega$ mass.
Assuming the SU(3) value of $1/3$ for the ratio of the 
$\omega$ to $\rho$
photo-couplings and adopting the form 
$\tilde\Pi_{\rho\omega}(s)=\tilde\Pi_{\rho\omega}(m_\omega^2)
+ (s - m_\omega^2)\tilde\Pi_{\rho\omega}'(m_\omega^2)$, the best fit to
the world data is
$\tilde\Pi_{\rho\omega}(m_\omega^2) = -3500\pm 300$
MeV$^2$ 
and $\tilde\Pi_{\rho\omega}'(m_\omega^2) = 0.03 \pm 0.04$~\cite{pionff97}.
We have assumed that 
$\tilde\Pi_{\rho\omega}(s)$ is real in the 
resonance region~\cite{debeau79}; relaxing this assumption yields
${\rm Im}\tilde\Pi_{\rho\omega}(m_\omega^2)= -300 \pm 300$ MeV$^2$, with
no change in the real part~\cite{pionff97}. 
Note that if 
finite width corrections, of importance for the $\rho$,
are included, the ratio of $\omega$ to $\rho$ 
photo-couplings decreases
by about 10\%~\cite{klingl97} and $\tilde\Pi_{\rho\omega}(s)$ becomes more
negative to the
same degree. Such corrections do not impact 
the sign of this quantity, which is determined by the pion
form factor in the vicinity of the $\omega$.
In contrast,
Enomoto and Tanabashi adopt the model of Ref.~\cite{gasser82} 
and make the above SU(3) assumption to find
a real, $s$-independent 
$\tilde \Pi_{\rho\omega}= -0.63\Gamma_\omega m_\omega 
\approx -4100$ MeV$^2$~\cite{ET96,footnote}. 
Using Eqs.(\ref{asym},
\ref{fun2}) we find
\be
re^{i\delta}\,e^{i\phi}= 
{ \langle \pi^+\pi^- \rho^- | {\cal H}^{\rm P} | B^- \rangle 
\over 
\langle \pi^+\pi^- \rho^- | {\cal H}^{\rm T} | B^- \rangle}
= 
{ \tilde\Pi_{\rho\omega} p_{\omega} + s_\omega p_\rho
\over
\tilde\Pi_{\rho\omega} t_{\omega} + s_\omega t_\rho} \;.
\ee
Recalling the definitions of Ref.~\cite{ET96}:
\be
{p_\omega \over t_\rho} \equiv r' e^{i(\delta_q + \phi)} \;, \quad
{t_\omega \over t_\rho} \equiv \alpha e^{i \delta_\alpha} \;, \quad
{p_\rho \over p_\omega} \equiv \beta e^{i \delta_\beta} \;,
\label{alphabeta}
\ee
one finds, to leading order in isospin violation, 
\be
re^{i\delta} = 
{r' e^{i\delta_q}\over s_\omega} \left\{
 \tilde\Pi_{\rho\omega} + \beta e^{i\delta_\beta}
\left( s_\omega 
- \tilde\Pi_{\rho\omega} \alpha e^{i\delta_\alpha} \right)
\right\} 
\;.
\label{thescoop}
\ee
Note that $\delta_\alpha,\delta_\beta,\hbox{and}\;\delta_q$ characterize
the remaining unknown strong phases. In the absence of isospin violation, 
as characterized here by $\rho^0$-$\omega$ mixing, at least one
of these phases would have to be non-zero in order to generate 
a rate asymmetry and hence CP violation. 
In the model of Bander {\it et al.}~\cite{bss79}, these phases are 
generated by putting the quarks in loops on their
mass-shell and thus are referred to as ``short-distance'' phases.

The resonant enhancement
of the CP-violating asymmetry is driven by 
$\tilde\Pi_{\rho\omega}/s_\omega$.
We stress that  
$\beta$ is essentially zero in $B^-\ra \rho^- \pi^+\pi^-$ because the 
gluon in the strong penguin diagram couples to an isoscalar
combination of $u\overline u$ and $d\overline d$. Thus it does not
couple to an isovector $\rho^0$ in the absence of isospin violation. 
Hence, $p_\rho$ as defined above is
non-zero {\em only} if electroweak penguin diagrams, naively suppressed by
$\alpha_{\rm em}/\alpha_{\rm s}$, or isospin violating effects in the
hadronic matrix elements which distinguish the 
$\rho^\pm$ and $\rho^0$ are included. Both effects were neglected in 
Ref.~\cite{ET96}. 
As $s\ra m_\omega^2$, the asymmetry is maximized if 
$|\chi|=|\tilde\Pi_{\rho\omega}|/m_\omega\Gamma_\omega\sim O(1)$ and
$\delta_q + \eta \sim \pm\pi/2$, where 
$\eta=-{\rm arg}\; s_\omega$. 
Using standard values for $m_\omega$ and $\Gamma_\omega$~\cite{pdg96} 
yields $|\chi|= .53$ and
$\eta= -\pi/2$. Note that $\delta_q$ as estimated in the factorization
approximation is $\rapp -\pi$ ~\cite{ET96}, so
that $\delta_q +\eta \rapp -3\pi/2$ at the $\omega$ mass. 
Thus, the participation of the $\omega$ resonance,  
with its narrow width, strongly enhances
the strong phase without the penalty of a severely small 
$|\chi|$. 

The CP-violating asymmetry from Eqs.~(\ref{asym},\ref{thescoop}), then, is 
determined by  
$\tilde\Pi_{\rho\omega}$, $m_\omega$, $\Gamma_\omega$, and the
``short distance'' parameters $\alpha$, $\delta_\alpha$, $\beta$,
$\delta_\beta$, $r'$, $\delta_q$, as well as  $\phi$, the weak phase
which we wish to determine. 
The crucial issue is therefore how well the latter 
parameters can be determined. 
As the 
sign of $\sin\phi$ is of unique significance, our particular focus
is on the short distance phase information 
required to extract it without ambiguity. 
The sign of the
CP-violating asymmetry in Eq.~(\ref{asym}) 
is determined by $\sin\delta$ and $\sin\phi$. 
The sign of $\sin\delta$ is in turn determined by 
$\cos\delta_q \,{\rm Im} \Omega + \sin\delta_q \,{\rm Re} \Omega$,
where $r\exp(i\delta)\equiv r'\Omega \exp(i\delta_q)/|s_\omega|^2$, noting
Eq.~(\ref{thescoop}). 
With $\chi\equiv \tilde \Pi_{\rho\omega}/(m_\omega \Gamma_\omega)$ 
as before, 
we find $|\chi| > \beta$, 
$\alpha\approx 1$, and ${\rm Im} \chi \ll {\rm Re}\chi$,
so that for $s\approx m_\omega^2$, 
\be
r \sin\delta \approx
\tilde \Pi_{\rho\omega}
\frac{r'}{|s_\omega|^2}
[(s- m_\omega^2) \sin\delta_q - m_\omega \Gamma_\omega \cos\delta_q]\;. 
\label{signsin}
\ee
The sign of $\sin\delta$ at $s=m_\omega^2$ is thus given
by $\tilde\Pi_{\rho\omega}$ and $\cos\delta_q$. 
The former is determined by $e^+e^-$ data, but 
what of $\cos\delta_q$? 
We will use the factorization approximation 
to compute $\cos\delta_q$ and thus its sign, 
yet the above asymmetry can also be used
to test its utility. 
As seen from Eq.~(\ref{signsin}),
the shape of the asymmetry with the invariant mass of the $\pi^+\pi^-$
pair is primarily of Breit-Wigner form, 
as dictated by the 
$1/|s_\omega|^2$ factor, although it will be 
``skewed'' if the coefficient of the $(s - m_\omega^2)$ term is
non-zero. Thus, the sign of the skew of the asymmetry is
driven by $\sin\delta_q$.
We show below that the empirical
$s$-dependence of $\tilde\Pi_{\rho\omega}(s)$ about $s=m_\omega^2$ 
and the magnitude
of its imaginary part at $s=m_\omega^2$ 
do not cloud this interpretation. Thus, we can extract 
$\tan\delta_q$. This does not 
fix the sign of $\cos \delta_q$, so that we use 
the factorization approximation to fix its sign and hence 
determine that of $\sin\phi$. 

We have chosen to study just one
of the channels considered in Ref.\cite{ET96}, namely $B^{\pm}
\ra \rho^{\pm} \rho^0(\omega)\ra \rho^{\pm} \pi^+ \pi^-$, as it is of
special character. In this case, the penguin amplitude to produce
a $\rho^0$ meson is zero but for isospin violation, so that $\beta$ is
nonzero only when 
electroweak penguins and isospin violating
effects which distinguish the $\rho^\pm$ and $\rho^0$ are included. 
As a consequence, we can associate the ``skew'' of the asymmetry with 
a single short distance quantity, $\sin\delta_q$, and ultimately test
the factorization approximation we apply. 

In principle, the short-distance parameters can be computed by using
the operator product expansion to construct an effective 
Hamiltonian at the $b$ quark scale, $\mu\approx m_b$. It is usually
written as a sum of tree and penguin
contributions, as anticipated in Eq.~(\ref{start}). Following 
Ref.~\cite{deshe95} we have, for example:
\be
{\cal H}^{\rm eff} = {4 G_F \over \sqrt{2}} 
          \{
V_{ub}V^*_{ud} 
          \sum_{i=1}^2 c_i(\mu) {\cal O}_i^{(u)} - 
V_{tb}V^*_{td} 
          \sum_{i=3}^{10} c_i(\mu) {\cal O}_i^{(u)} 
\} + {\rm H.C.} \;,
\label{effham}
\ee
with
${\cal O}_1^{(u)} =
{\overline d}_{L\alpha} \gamma^\mu u_{L\beta}{\overline u}_{L\beta}
\gamma_{\mu} b_{L\alpha}$ and 
${\cal O}_2^{(u)} =
{\overline d}_{L} \gamma^\mu u_{L}{\overline u}_{L}
\gamma_{\mu} b_{L}$~\cite{deshe95}.
Ten four-quark operators characterize the effective Hamilitonian; 
$i=3,\cdots, 6$ label the strong penguin operators, 
whereas $i=7,\cdots, 10$ label the electroweak penguins. 
The Wilson
coefficients $c_i$ are known through
next-to-leading logarithmic order~\cite{deshe95}, 
yet consistency to one-loop-order requires that the matrix elements be
renormalized to one-loop-order as well~\cite{flrev96}. This renormalization
procedure results in {\it effective} Wilson coefficients 
$c_i'$ which multiply the
matrix elements of the given operators at tree level. 
The effective Wilson coefficients develop an imaginary part 
if the quarks in loops 
are set on their mass-shell; to compute them, 
we use the analytic expressions of Ref.~\cite{kramer94}
with a charm quark mass of 
$m_c=1.35$ GeV. 
There is some sensitivity to $k^2$, the invariant mass of the 
exchanged boson, 
and thus we use two values of 
$k^2$, $k^2/m_b^2 = .3, .5$,  covering the expected 
``physical'' range~\cite{buras95}. 

We now turn to the evaluation of the matrix elements of this effective
Hamiltonian. We use the factorization 
approximation, so that 
$\langle \rho_I^0 \rho^- | 
{\overline d}_L\gamma^\mu u_L {\overline u}_L \gamma_\mu b_L | B^-\rangle
= \langle \rho^- | 
{\overline d}_L\gamma^\mu u_L | 0 \rangle 
\langle \rho_I^0 | {\overline u}_L \gamma_\mu b_L | B^-\rangle
+ {(1 / N_c)} 
\langle \rho_I^0 | 
{\overline u}_L\gamma^\mu u_L | 0 \rangle 
\langle \rho^- | {\overline d}_L \gamma_\mu b_L | B^-\rangle$. 
$N_c$ is the number of colors, but here it is treated as a phenomenological
parameter.
Fits to measured branching ratios in $B\ra D^\ast X$ decays indicate that
the empirical value of the ratio $(c_1'+c_2'/N_c)/(c_2' + c_1'/N_c)$ 
is bounded by $N_c=2$ and $N_c=3$~\cite{rodrig96}. 
Large $N_c$ arguments 
justify the factorization approximation~\cite{buras86}, yet 
$N_c=\infty$ 
yields a ratio of the wrong sign. 
Thus, we use $N_c=2,3$ as an empirically constrained gauge of the
uncertainties inherent in the factorization approximation in what follows. 

In the preceding paragraph,
$\rho_I^0$ (and $\omega_I$) denote isospin-pure states, for
$\tilde\Pi_{\rho\omega}$ 
characterizes isospin violation in the $\rho^0$ and
$\omega$. Consistency requires
that we compute all other sources of isospin violation to the
same order. In particular, we need to estimate how isospin violation
distinguishes $\rho^\pm$ from $\rho_I^0$ and 
$\omega_I$ in the hadronic matrix elements.
This additional source of isospin breaking can 
be parametrized via 
\be
{ 
\langle \rho^- | {\overline d}_L\gamma^\mu u_L 
 | 0 \rangle 
\langle \rho_I^0 | {\overline u}_L\gamma_\mu b_L | B^- \rangle 
\over
\langle \rho_I^0 | {\overline u}_L\gamma^\mu u_L | 0 \rangle
\langle \rho^- | {\overline d}_L\gamma_\mu b_L | B^- \rangle }
\equiv 1 + \tilde\varepsilon \;.
\ee
We use the model of Ref.~\cite{bsw87}
to evaluate the $B^- \ra \rho^-,\rho_I^0$ transition
form factors and find that $|\tilde\varepsilon|$ is no larger than
$0.01$. The resultant $\alpha$, $\beta$,
etc., as per Eq.~(\ref{alphabeta}), are, 
defining $\xi\equiv 4((c_3'+ c_4')(1 + 1/N_c) + c_5' + c_6'/N_c) + 
(c_9'+ c_{10}')(1 + 1/N_c)$ and assuming $t$ quark dominance, 
$\alpha \exp(i\delta_\alpha) = 1$, 
\begin{mathletters}
\label{result}
 \begin{eqnarray}
\beta e^{i\delta_\beta} =
\frac{3}{\xi}(c_9'+ c_{10}')(1 + \frac{1}{N_c})  
+ \frac{2\tilde\varepsilon}{\xi} 
(c_4' + \frac{c_3'}{N_c} + c_{10}' + \frac{c_9'}{N_c})
( 1 - \frac{3}{\xi}(c_9'+ c_{10}')(1 + \frac{1}{N_c}) ) \;,
\\
r' e^{i(\delta_q + \phi)} =- 
\left\{
{\xi + 2\tilde\varepsilon (c_3'/N_c + c_4' + c_9'/N_c + c_{10}') 
\over 2(c_1' + c_2')(1 + 1/N_c)}
-
{(c_1'/N_c + c_2')\tilde\varepsilon \xi
\over 2((c_1' + c_2')(1 + 1/N_c))^2}
\right\}{V_{tb}V_{td}^* \over V_{ub} V_{ud}^* } \;.
\end{eqnarray}
\end{mathletters} 
We use 
the parameters of Ref.~\cite{ET96} to evaluate $V_{ij}$.
The CP-violating 
asymmetry, which follows from Eqs.~(\ref{asym},\ref{alphabeta},
\ref{thescoop},\ref{result}), is shown in Fig.~1a as a function
of the invariant mass of the 
$\pi^+\pi^-$ pair. 
The asymmetry is no less
than 20\% at the $\omega$ mass. The asymmetry really is driven by
$\rho^0$-$\omega$ interference as the same asymmetries, now
with ${\rm Im}(c_i')=0$, are shown in Fig.~1b. 
In the absence of the short-distance phases,
the asymmetry is symmetric about the $\omega$ mass. 
Figure 1c shows the sensitivity of the 
$N_c=2$, $k^2/m_b^2=0.5$ asymmetry to the error in 
$\tilde\Pi_{\rho\omega}(m_\omega^2)$, and 
Fig.~1d shows the sensitivity of the 
same asymmetry to the allowed
$\tilde\Pi_{\rho\omega}'$  and 
${\rm Im}(\tilde\Pi_{\rho\omega})$ contributions~\cite{pionff97}.
Both of the latter generate a slight ``skew'' to the
shape of the asymmetry about the $\omega$ mass, yet these effects are
sufficiently small for it to be meaningfully associated with 
the short distance parameters. 
Our careful computation of the effects which would
generate a non-zero $\beta$ allows us to conclude that 
$\beta$, which ranges from $0.12 - 0.18$, 
is indeed smaller than $|\chi|\sim 0.53$, so that the skew of the
asymmetry constrains $\sin\delta_q$. 
{\em Indeed, a measurement of the shape of the asymmetry 
constrains whether any long-distance phase accrues in the breaking
of the factorization approximation} -- i.e., through 
$q\overline q$ pair creation in the full matrix elements.

We have computed the CP-violating asymmetry in 
$B^\pm \ra \rho^\pm \rho^0(\omega) \ra \rho^\pm \pi^+\pi^-$ decay
and have found that the asymmetry, which is greatly enhanced
through $\rho^0$-$\omega$ interference, is significantly constrained
through $e^+e^-\ra \pi^+\pi^-$ data in the $\rho^0(\omega)$ interference
region. Indeed, the magnitude of the asymmetry would be
preserved even if $\delta_q$,
the phase arising from the effective Wilson coefficients, 
were zero, though its {\it sign} does depend on 
$\cos\delta_q$. In the factorization approximation, 
$\cos\delta_q < 0$ for any $N_c>0$ and $k^2/m_b^2$, as 
the magnitude of $\tilde\epsilon$ is set by that of isospin violation. 
Thus, for the decay of interest, the factorization approximation
fails to predict the sign of $\cos\delta_q$ only if it is badly 
wrong. Fortunately, moreover, the measurement itself provides a 
significant test of the factorization approximation, for 
$\tan\delta_q$ can be extracted as well. 
This is a much more germane test of its utility
than that afforded by empirical branching ratios. 
Thus, 
we are led to conclude that the CP-violating asymmetry in the above
channel is large and robust with respect to the known strong
phase uncertainties, admitting the extraction of the weak phase $\phi$,
or specifically $-\alpha$~\cite{pdg96}, from this channel.  

S.G. thanks A. Kagan, W. Korsch, and G. Valencia for 
helpful comments and references, and the University of Adelaide for
hospitality during a visit when this work began. 
This work was supported by the DOE under DE-FG02-96ER40989 (SG and HOC)
and by the Australian Research Council.


\begin{figure}
\caption{The CP-violating asymmetry, Eq.~[\protect{\ref{asym}}], in percent,
plotted versus the invariant mass $q$ of the $\pi^+\pi^-$ pair in MeV for
[$N_c$, $k^2/m_b^2$].
a) The asymmetries with $\tilde\Pi_{\rho\omega}=-3500\;{\rm MeV}^2$ and 
$\tilde\varepsilon=-0.005$ are shown for [2, 0.5] (solid line), 
[3, 0.5] (long-dashed), [2, 0.3] (dashed), and [3, 0.3] (dot-dashed). 
b) The asymmetries of a) are shown with ${\rm Im}(c_i')=0$. 
c) The [2, 0.5] asymmetry of a) is shown (solid),
with $\tilde\Pi_{\rho\omega}=-3200\;{\rm MeV}^2$ (long-dashed) and with
$\tilde\Pi_{\rho\omega}=-3800\;{\rm MeV}^2$ (dashed). 
d) The [2, 0.5] asymmetry of a) is shown (solid),
with $\tilde\Pi_{\rho\omega}'=0.027$ (long-dashed) and with
${\rm Im}(\tilde\Pi_{\rho\omega})=-300\;{\rm MeV}^2$ (dashed). }
 \label{figone}
\end{figure}

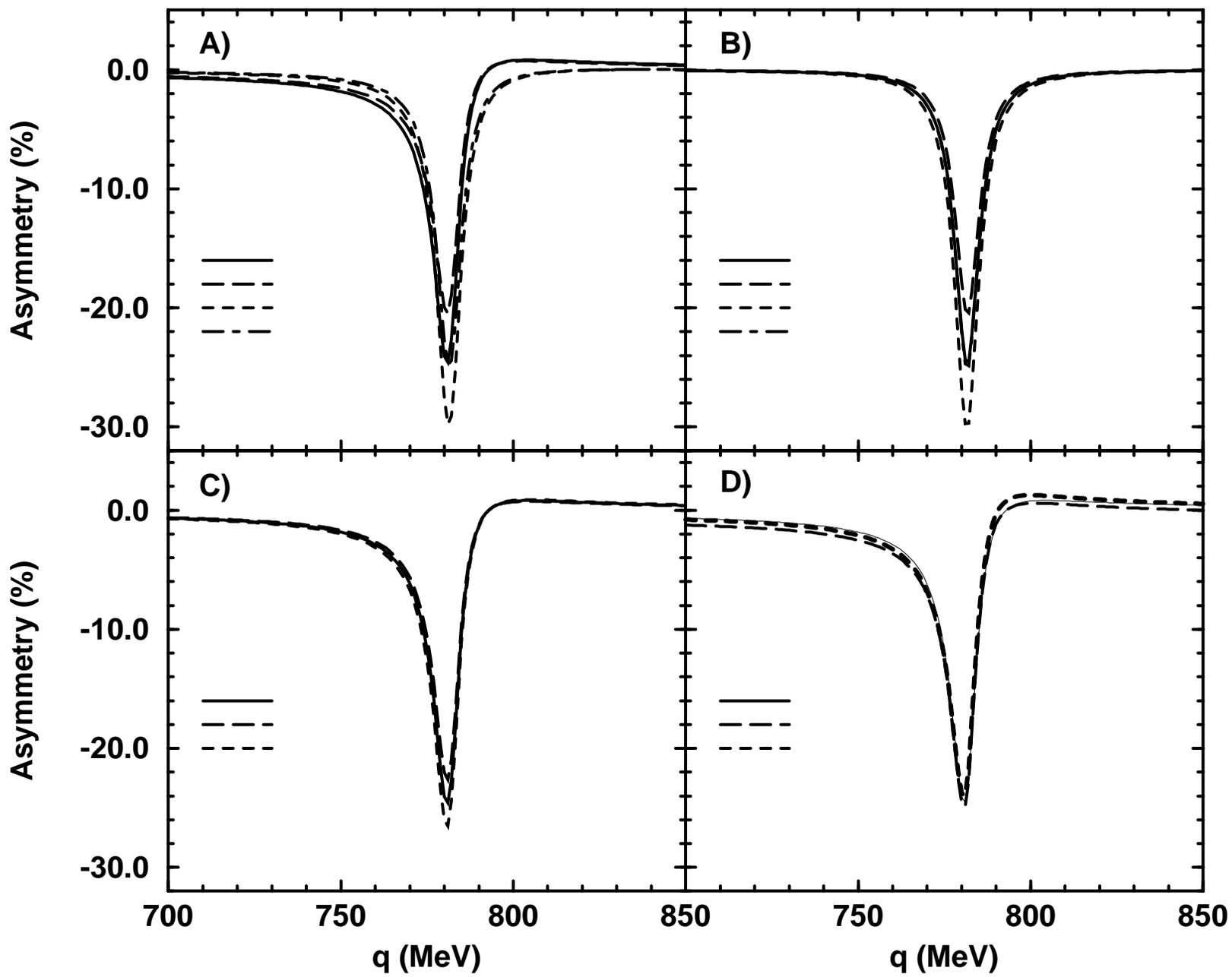
\end{document}